\documentclass[12pt]{article}
\usepackage{nsf-minaas,apjfonts,graphicx,natbib,enumitem}
\usepackage[font=small,labelfont=bf,labelsep=period]{caption}
\PassOptionsToPackage{hyphens}{url}\usepackage{hyperref}

\hypersetup{
  colorlinks = true, 
  urlcolor = [rgb]{0.0,0.4,0.8}, 
  linkcolor    = [rgb]{0.7,0.0,0.3}, 
  citecolor   = [rgb]{0.7,0.0,0.3}, 
  breaklinks = true
}

\setlist[enumerate]{font=\bfseries}

\bibpunct{(}{)}{;}{a}{}{,}
\begin{document}

\begin{center}
\Large{\textbf{HST/JWST Long-Term Monitoring Working Group \\ Final Report}}\\

\vskip 0.1in 
\large
Saurabh W.~Jha\footnote{Rutgers University} (co-Chair), 
Dana I.~Casetti-Dinescu\footnote{Southern Connecticut State University} (co-Chair), 
Gary M.~Bernstein\footnote{University of Pennsylvania}, \\
Matthew J.~Hayes\footnote{Stockholm University}, 
Lidia M.~Oskinova\footnote{Potsdam University}, 
Andrew B.~Pace\footnote{Carnegie Mellon University}, 
Robert M.~Quimby\footnote{San Diego State University}, \\
Megan Reiter\footnote{Rice University},
Armin Rest\footnote{Space Telescope Science Institute},
Adam G.~Riess\footnote{Johns Hopkins University and Space Telescope Science Institute}, 
David J.~Sand\footnote{University of Arizona}, and 
Daniel R.~Weisz\footnote{University of California, Berkeley} \\

\vskip 0.1in
STScI Facilitators: Laura Watkins and I.~Neill Reid \\

\vskip 0.1in
May 17, 2024
\end{center}

\normalsize
\section*{Abstract}

\vskip -0.2in 
The Astro2020 Decadal Survey recognizes time-domain astronomy as a key science area over the next decade and beyond. With over 30 years of HST data and the potential for 20 years of JWST operations, these flagship observatories offer an unparalleled prospect for a half-century of space-based observations in the time domain. To take best advantage of this opportunity, STScI charged a working group to solicit community input and formulate strategies to maximize the science return in time-domain astronomy from these two platforms. Here, the HST/JWST Long-Term Monitoring Working Group reports on the input we received and presents our recommendations to enhance the scientific return for time-domain astronomy from HST and JWST. We suggest changes in policies to enable and prioritize long-term science programs of high scientific value. As charged, we also develop recommendations based on community input for a JWST Director's Discretionary Time program to observe high-redshift transients.

\clearpage

\section{Charge}

Time domain astronomy was highlighted as a key science area for the 2020s in the Astro2020 Decadal Review \citep{NAS:2021}. Both HST and JWST have the potential to make significant scientific contributions in probing the variable universe at moderate and longer timescales. HST has the advantage of a 30-year legacy of observations spanning the broadest range of celestial sources; JWST has the prospect of a 20-year lifetime, with unparalleled sensitivity at near- and mid-infrared wavelengths also opening up opportunities in the high redshift Universe. Consequently, these two platforms offer unprecedented advantages to study photometric, spectroscopic, and astrometric variability. Furthermore, concurrent operations of these two observatories can enhance the science return through broad wavelength coverage and cross-calibration. 

In early 2023, the STScI leadership developed a plan for a working group to provide guidance on optimal strategies for maximizing the scientific return from HST and JWST time-domain observations. The Long-Term Monitoring Working Group (LTMWG) was formed in March 2023 and charged with: 1) soliciting community input on key science areas, 2) identifying science themes to be prioritized, 3) recommending new policies and provide advice on existing policies and resources needed to enable long-term science, and 4) developing a specific concept for an observing program that will utilize JWST’s imaging and spectroscopic capabilities to probe transient phenomena at high redshift, with the goal of starting implementation of the program in JWST Cycle 2. The members of the working group are the authors of this report and we were joined \emph{ex officio} by I.~Neill Reid (STScI Associate Director for Science) and Laura Watkins (STScI Deputy Head, Science Mission Office).

\section{Process and Community Input}
The LTMWG met approximately weekly throughout the northern spring and summer of 2023 and developed a strategy to solicit community input. We identified two major topics to focus our requests to the community and subsequent deliberation:
\begin{itemize}
\item {\bf Long time baseline science opportunities}, where long time baselines refer to those that are not easily accommodated in the standard proposal process. Observations could include, but are not limited to, photometric or spectroscopic variability, and astrometric motions. 

\item {\bf JWST DDT for high-redshift transients}, meaning a specific concept for a Director’s Discretionary Time observing program that will use JWST’s imaging and spectroscopic capabilities to probe transient phenomena at high redshift, with a goal of starting implementation of the program in JWST Cycle 2. What are the science cases that should be prioritized for such a program?
\end{itemize}

The LTMWG sent a call for community input via an STScI email on June 15, 2023 (\autoref{sec:call_for_input}). We requested input in two forms: an online survey (\autoref{sec:community_survey}) with a set of questions to be addressed submitted anonymously online, and/or a 1--2 page white paper to be emailed to STScI. The deadline for responses was set to September 8, 2023. Our working group also held a virtual Town Hall with approximately 100 attendees on August 17, 2023, where we presented an overview of our activities and reiterated our request for community input. A video recording of the Town Hall, our request for community input, and our working group charter were \href{https://outerspace.stsci.edu/display/HPR/Long-term+variability+monitoring+strategies+for+HST+and+JWST}{posted online}.

We received a robust community response comprising 91 survey submissions and 38 white-paper contributions, with many authors. In our requests, we noted that community input would guide our working group's discussions and recommendations, but we would not directly share the submissions publicly. Our motivation was to allow community members to freely share ideas that might otherwise be proprietary (e.g., in planned HST or JWST proposals). We encourage astronomers who submitted material to our working group to publicize their contributions as they wish.

The consensus from the community is that \emph{existing policies are not adequate in enabling long-term science}: nearly all of the white paper contributions addressing long-term science wanted to see the current policies changed. 

\begin{figure}
    \centering
    \includegraphics[width=0.6\linewidth]{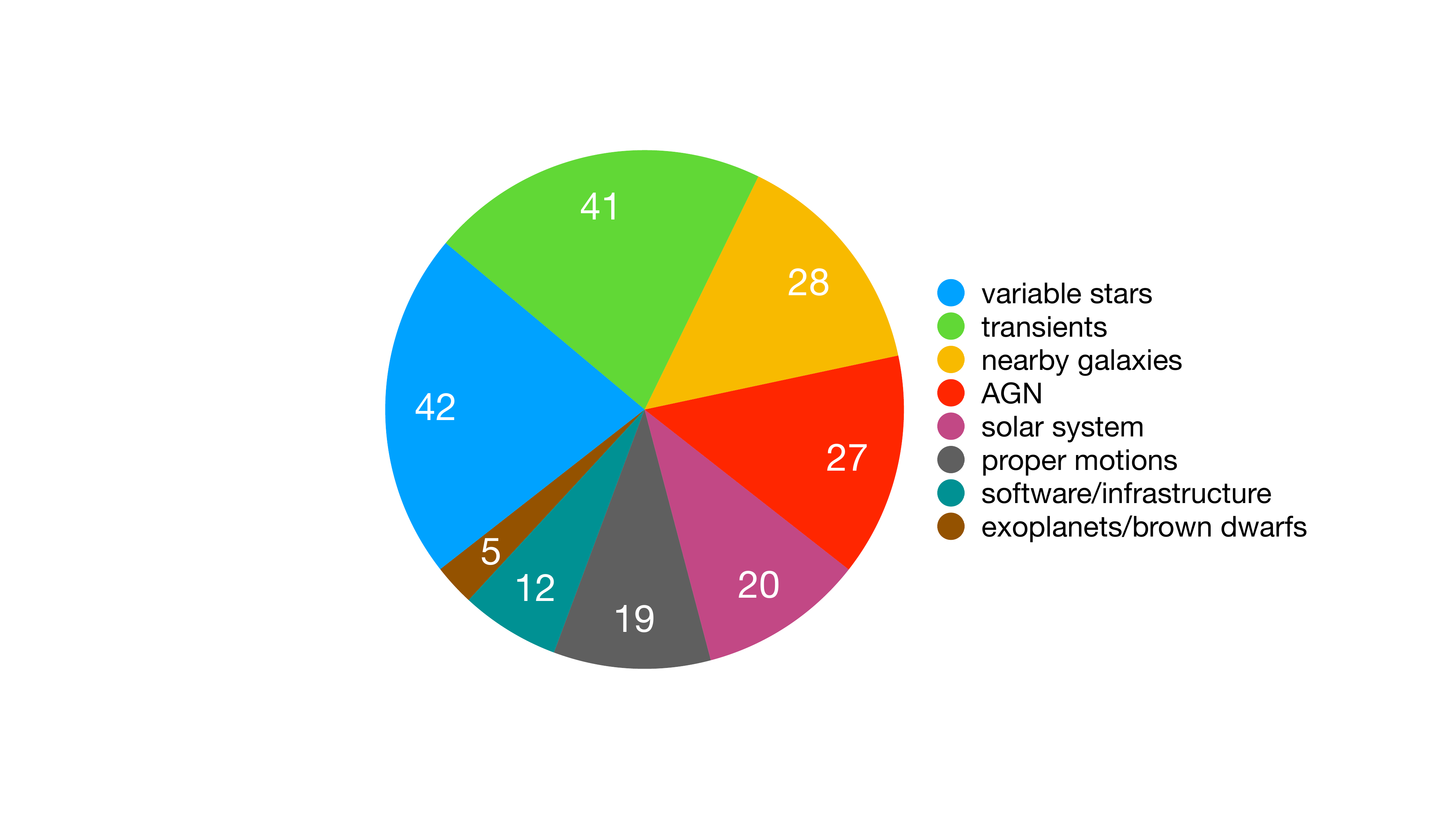}
    \caption{Long time-baseline topics of interest prioritized in community survey responses.}
    \label{fig:topics}
\end{figure}

The long time-baseline science topics highlighted by the community span the range from our solar system to high-redshift galaxies and AGN, underscoring the wide impact of this mode of science (\autoref{fig:topics}). Specifically, the science areas prioritized in the survey responses were variable stars (42), transients (41), nearby galaxies (28), AGN (27), solar system (20), proper motions (19), software/infrastructure (12), and exoplanets/brown dwarfs (5). We note that exoplanet research is also being considered by a dedicated working group \citep{Redfield:2024}. We thank the community for their input and interest and hope that we have done our best to bring to focus their concerns and requirements. 

\section{Long Term Variability and Monitoring Programs}

\begin{figure}[!b]
    \centering
    \includegraphics[width=0.9\linewidth]{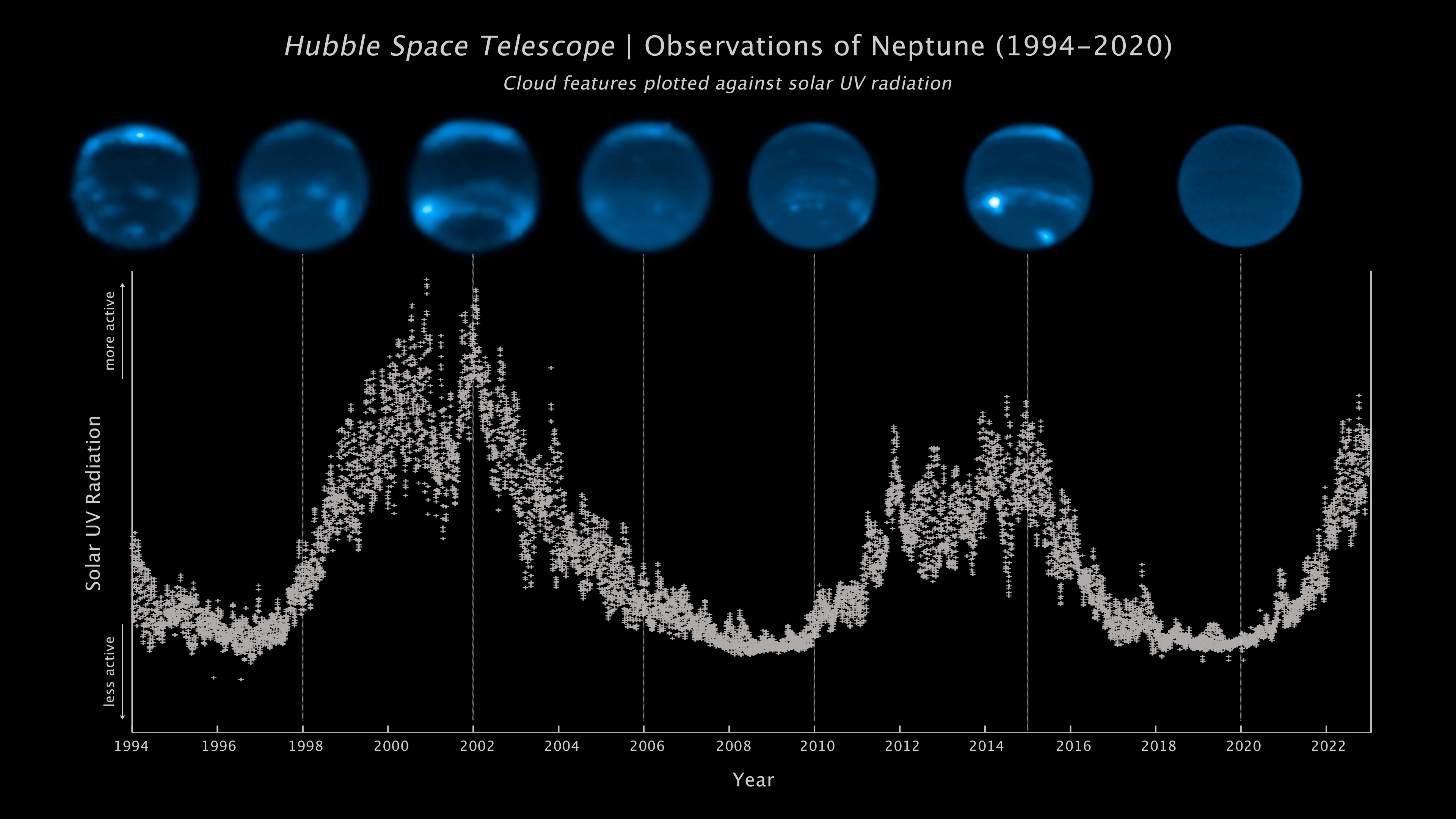}
    \caption{Nearly three decades of HST imaging of Neptune shows the correlation between cloud features and solar activity \citep{Chavez:2023}. Image credit: NASA, ESA, LASP, Erandi Chavez (UC Berkeley), Imke de Pater (UC Berkeley), Joseph DePasquale (STScI), from \href{https://hubblesite.org/contents/news-releases/2023/news-2023-019}{HubbleSite news release 2023-091}. \label{fig:neptune}}
\end{figure}

\subsection{Scientific Motivation}
The science topics that would benefit most from revised policies are those that need two or more epochs of observations widely spaced in time and those that need regular monitoring (e.g., yearly) over a time baseline longer than can be accommodated in a single proposal. Programs in this latter category typically need only a modest amount of observing time per epoch, repeated regularly (e.g., yearly).

In the first group, proper motions of distant targets such as Local Group galaxies and nearby galaxies, or stars in dense cluster environments, are certain to benefit from 20--40 year baselines \citep[e.g.][]{Kallivayalil:2013, Sohn:2013, Bellini:2014}. Currently, there are \emph{no} platforms that offer the depth, resolution, and time baselines of HST and JWST in such studies. Archival HST data have already proven to be extremely useful for the proper motion studies; for instance, \citet{Libralato:2021} used two epoch of HST data separated by $\sim$3 years to address the 2D kinematics of massive stars near the Galactic Center. Recently, \citet{Haberle:2024} used HST data spanning over a $\sim$20 yr baseline to study proper motions of stars in the most massive Galactic globular cluster, $\omega$ Cen, showing evidence for the presence of an intermediate mass black hole with $M >$ 8200 M$_\odot$ \citep{Haberle:2024b}. The HST observations allowed for detailed investigation of the color-magnitude diagram to separate the multiple stellar populations of the cluster, and the long time baseline improved upon previous proper motion constraints on the black hole \citep{Anderson:2010}. 

Other such examples are numerous, including studies of the large-scale rotation of the Large Magellanic Cloud based on full three-dimensional velocity measurements \citep{Marel:2014} and proper motions of dwarf spheroidal galaxies, such as Fornax \citep{Piatek:2007}. Long time baseline observations also enable stellar astrophysics, e.g., measuring the kicks of neutron stars unbound by supernovae in binaries \citep{Chrimes:2023}, and even extend to the distant Universe, e.g., discovering high-redshift AGN via photometric variability \citep{Hayes:2024}. Furthermore, long-term programs have huge potential for serendipitous discoveries, e.g., \citet{Sahu:2022} detected an isolated stellar-mass black hole using astrometric microlensing supported by a $\sim$6 yr span of HST data.  

The careful astrometric calibration of HST (which will also help calibrate JWST) ensures a degree of precision and accuracy of such measurements that no other existing platform can offer for distant targets. These measurements are critical in constraining --- via 3D dynamics --- ``near-field'' cosmology, which shows signs of tension with large-scale cosmological results \citep{Simon:2019}. More recently, measurements of distant Milky Way satellites as well as Andromeda satellites \citep{Bennet:2023, Casetti-Dinescu:2022, Sohn:2020} are showing great improvements in precision that will strengthen cosmological implications from the Local Group. Specifically, the alignment of the orbital angular momenta of satellites in both the Milky Way and Andromeda systems is now on a more sure footing than previously, underlying the tension with cosmological predictions \citep{Pawlowski:2021, Taibi:2024}.

In the second group, monitoring programs study a wide range of science, such as giant planets in our solar system, star formation, stellar evolution, stellar death, Pop III stars, and AGN. These will also benefit from a coherent, judicious strategy for long-term programs. A beautiful example is the work of \citet{Chavez:2023}, shown in \autoref{fig:neptune}, using HST imaging observations of Neptune spanning nearly 30 years to connect the planet's cloud cover to solar UV radiation, rather than ``seasonal'' effects that operate on a longer timescale. 

Other examples of the science impact of long-term monitoring include the multi-epoch HST observations to determine the expansion center and dynamical age of the supernova remnant Cassiopeia~A \citep{Thorstensen:2001}, and studies of SN\,1987A that will continue into the JWST era \citep{Kangas:2023}. The monitoring of evolved stars and their associated nebulae during the fastest stages of their evolution, such as $\eta$\,Car outflow \citep{Mehner:2011} or a planetary nebula N11 in the Large Magellanic Cloud \citep{Hamann:2003} is necessary to understand these rare but important objects. Consistent monitoring is also invaluable for studies of outflows from forming stars \citep{Krist:2008, Rich:2020} and jets from accreting compact objects \citep{Perlman:2003}.    

\subsection{Recommendations}

Based on the response we received, there is strong community interest in long-term programs with HST and JWST spanning the widest range of science topics. The current proposal process does not adequately enable this science for observations that need longer than a three-cycle time baseline.

\begin{enumerate}

\item \textbf{Our primary recommendation is that long-term science should be enabled and prioritized for HST and JWST.}

\end{enumerate}

Implementation of this recommendation will require input and discussion from STScI and other stakeholders. Here, we suggest changes to policies and processes aimed at this goal, with the understanding that details may need further consideration at STScI. We also expect ongoing experimentation, assessment, and as needed, modification of the implementation in furtherance of this primary goal.

Some long-term science might be relatively easily accommodated with an increase in the time baseline allowed for multicycle programs.

\begin{enumerate}[resume]
\item \textbf{We recommend consideration of whether the three cycle limit for multicycle observations in a single proposal should be extended.} The limit should be determined with a scientific rationale. 
\end{enumerate}

Hereafter, we use “long-term” to refer to programs requiring observations beyond the multicycle window for single proposals, whether that is the current policy of three cycles or some modified length.

We can categorize most long-term programs in two groups: 1.~those that require widely-separated epochs of observations (e.g., proper motion measurements); or 2.~those that require regular monitoring over a long time baseline. Changes to policies and procedures should aim to enable both types of programs. 

\begin{enumerate}[resume]
\item \textbf{We recommend proposers identify Long-Term Programs through an APT checkbox.} Proposers should be instructed that checking this box requires additional information to be supplied in the “Special Requirements” section of the proposal, including details of the future observations.

\item \textbf{We recommend that Long-Term Programs be assigned to a separate panel in the proposal review.} This panel will require expertise in a wide range of science, but can be attuned to the specific issues of Long-Term Programs. In particular this panel should be instructed how to judge the proposed science and technical aspects of the overall program, rather than just the near-term observations. Even if a separate panel for Long-Term Programs is not feasible, it is critical that reviewers be given explicit instructions on how to judge these programs to allow consideration of the overall program and not disadvantage them compared to shorter-timescale programs.

\item \textbf{We recommend consideration for a pool of observing time for Long-Term Programs.} This pool can be used for subsequent epochs in approved programs.

\item \textbf{We recommend establishing a mechanism to review progress for Long-Term Programs and suggest changes if necessary.} This would allow monitoring programs to proceed without repeatedly requiring panel review, but also allow programs to be stopped or modified if circumstances change.
\end{enumerate}

We note that these recommendations raise a number of issues for further consideration, including coordination of Long-Term Programs with other observatories, accounting for the time allocated to future epochs of observations, proprietary periods for Long-Term Programs, funding allocations for Long-Term Programs, follow-up of discoveries from monitoring programs, and changing proposers, multiple teams, or duplicated targets in Long-Term programs. The new “Archival+GO” category could be broadened (including to non-US investigators) to provide an avenue for some of these programs to connect past and future epochs.

Separately from GO Long-Term Programs that can go through the regular proposal process, we note we are at the beginning of an expected long life for JWST and this provides a key opportunity for a concerted effort (larger than can be expected for individual proposals) to acquire early data for long-term studies:

\begin{enumerate}[resume]

\item \textbf{To establish JWST legacy observations with the longest time baseline, we recommend a community process to identify highest-priority fields to be observed with JWST as soon as feasible.} As an example, we anticipate observations could be for first-epoch astrometry of selected fields (e.g., nearby galaxies), with significant ancillary science, that could then be followed up with JWST, or future facilities (e.g,. Habitable Worlds Observatory), on timescales of decades. We recommend either Director’s Discretionary Time or a call for Treasury proposals to obtain these early data.

Contemporaneous HST observations may allow tying JWST astrometry to past HST astrometry. The astrometric calibration of JWST should be monitored and prioritized throughout its operation. These data should have legacy value for future observatories beyond HST and JWST.

\end{enumerate}

\section{JWST DD Program for High-Redshift Transients}

\subsection{Scientific Motivation}

Our working group charter also included direction to ``develop a specific concept for an observing program that will utilize JWST’s imaging and spectroscopic capabilities to probe transient phenomena at high redshift, with the goal of starting implementation of the program in JWST Cycle 2.''

For our purposes, we interpret ``high redshift'' in this charge to mean using JWST to target transients that are more distant than HST has been able to observe.

A key mission success science goal for JWST is to observe the first generation of stars and galaxies. \textbf{Discovering a Pop III supernova}, marking the death of one of the first stars, would achieve this, and \textit{is something only JWST can do.} However, there are major uncertainties in the rates and properties of Pop III SN so that we cannot be sure we will be able to survey enough volume to detect one, nor do we know exactly what we are looking for. 

Theory predicts that the metal-free cooling of primordial gas should lead to the preferential formation of high-mass stars $>$100 M$_\odot$ \citep{Klessen:2023}. Stars in the mass range 140--260 M$_\odot$ are thought to explode as ``pair-instability supernovae'' \citep[PISN;][]{Woosley:2002} and should have distinguishing properties from the explosion of lower mass stars \citep[e.g.,][]{Kasen:2011,Whalen:2013,Gilmer:2017}. Discovery of an enhanced rate of high-mass stellar explosions at high redshift could be promising evidence that we have observed the first generation of stars and guide theoretical development.

\begin{figure}
    \centering
    \includegraphics[width=0.75\linewidth]{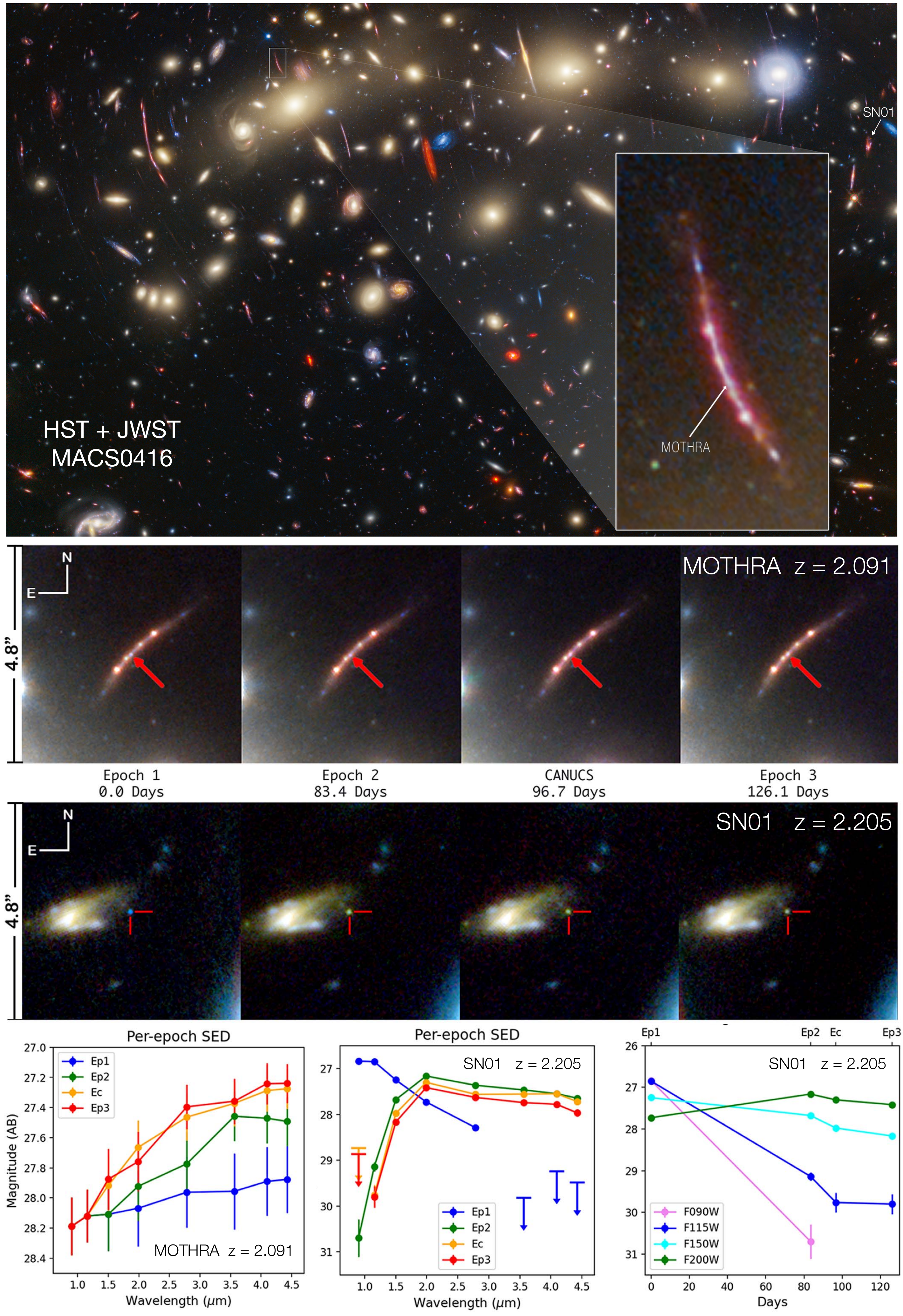}
    \caption{JWST discovery of transients in the Hubble Frontier Field MACS0416, adapted from \citet{Yan:2023} and \href{https://webbtelescope.org/contents/news-releases/2023/news-2023-146}{JWST news release 2023-146}. The panels show the highly gravitationally magnified star ``Mothra'' and the supernova SN01, both at $z > 2$. JWST is uniquely capable of measuring the SED and light curves (lower panels) of these transients. Upper panel image credit: NASA, ESA, CSA, STScI, Jose M. Diego (IFCA), Jordan C. J. D'Silva (UWA), Anton M. Koekemoer (STScI), Jake Summers (ASU), Rogier Windhorst (ASU), Haojing Yan (University of Missouri), and Joseph DePasquale (STScI), including HST data from programs 12459 (Postman), 13496 (Lotz), and 13386 (Rodney) as well as JWST data from programs 1176 and 2738 (Windhorst). \label{fig:pearls}}
\end{figure}

While such an observation is challenging due to low predicted rates \citep[e.g.,][]{Tanaka:2013,Regos:2020,Moriya:2022,Venditti:2024}, we aim to design a program that is capable of detecting Pop III SNe (including through their “shock breakout” emission) if the rate is high enough (or we are lucky enough). This aspirational idea sets requirements for depth, wavelength coverage, and cadence. In aiming for this lofty target, our recommended program will discover many high-redshift transients also of high scientific priority, including:

\begin{itemize}

\item discovery of tens of SN Ia from $z > 2$ out to $z \simeq 5$ (and further if they exist) potentially enabling a significant extension of the SN Ia Hubble diagram, and using the measured rates, providing unique constraints on the progenitors of SN Ia, including probing SN Ia from the first generations of stars.

\item many $z > 6$ core-collapse supernovae (rates of CCSN and subtypes can track changes in the IMF; light curves and spectra can probe changing metallicity, the role of binary interactions in massive star evolution, and the formation rates of binary compact objects, neutron stars and black holes).

\item variable AGN at $z > 7$ can help determine the seeding mechanism for supermassive black holes. Photometric variability of intermediate luminosity galaxies at redshifts beyond the ‘quasar epoch’ ($z >$ 2--4) will efficiently constrain the abundance of AGN at early times \citep[e.g.,][]{Hayes:2024}. This becomes especially critical at $z \gtrsim 7$, where the comoving number density of AGN will constrain the seeding mechanisms of the first supermassive black holes. For example, number densities in excess of a few $\times 10^{-3}$ cMpc$^{-3}$ at $z \gtrsim 7$ begin to point directly towards early formation by the collapse of isolated Population III stars \citep[][and depending somewhat on the precise redshift]{Singh:2023}, which will be analyzed in concert with the explosive transient aspects of this monitoring campaign.

\item rarer transients will also be discovered, including superluminous supernovae (SLSN) out to $z \simeq 10$ and tidal disruption events (TDEs) out to $z \simeq 5$. Detections at even higher redshifts are possible if the rates prove favorable.

\end{itemize}

JWST has already shown itself adept at discovering high-redshift transients in repeat observations. As one example, \citet{Yan:2023} present transients discovered in the Hubble Frontier Field MACS0416, observed with JWST as part of the PEARLS survey (\autoref{fig:pearls}). \citet{Yan:2023} emphasize ``the power of JWST in the study of the transient IR sky,'' with its combination of depth, FOV, and wavelength coverage far surpassing all previous observational capabilities in discovering and characterizing these distant transients, and they mark the beginning of ``a new era of IR time-domain science.''

\subsection{Recommendations}

We recommend three kinds of fields for observation, with a rough (and preliminary) suggestion of how much time to spend on each. Not only will the survey be transformative for high-redshift transients, it will also provide a wealth of ancillary science, ranging from solar system objects to high-redshift static sky deep fields.

\begin{enumerate}[resume]

\item \textbf{Our highest priority recommendation is cadenced NIRCam imaging of fields in the JWST CVZ.} The CVZ is the best location for long-term transient science, allowing observation (including followup) at any time. Fields near the North Ecliptic Pole (NEP) can leverage existing observations (e.g., the NEP Time Domain Field from the PEARLS program), but low-extinction, uncrowded fields near the Southern Ecliptic Pole (SEP) should also be considered for opportunities with current and future ground-based southern hemisphere facilities. These fields will become new benchmark deep fields for multiwavelength study.

This imaging should reach roughly 28th mag (5$\sigma$ depth) in at least three or four filters (e.g., F150W, F200W, F277W, F444W) for each epoch. The full area should be imaged at least twice a year. A subset of the area, perhaps prioritized by interesting active transients, should be imaged at least four times a year. Such a dynamic tier system utilizes the wide-field observations to obtain the widest field possible with a cadence suitable for the discovery and follow-up of high-z transients, while at the same time discovering a significant number of transients with shorter timescales in the observer frame. In each year, different high-cadence fields can be selected that optimize the scientific yield.

While most of the high-cadence fields can be dynamically prioritized, two or four static pointings should be observed consistently with a cadence of four times per year (e.g., with 90 degree rotations of NIRCam to cover the same FOV), perhaps with increased depth and filter coverage per epoch for these pointings. 

We recommend approximately 80 hours per year for these imaging observations, extending for at least three years. This will set the total area (we anticipate it approaching 0.1 deg$^2$). The two or four static high-cadence pointings should continue to be observed for five years. In sum, we expect this survey to require approximately 260 hours (80 hours $\times$ 3 years = 240 hours + 10 additional hours $\times$ 2 more years for the static high-cadence pointings).

The approach outlined here should be modified as needed: the goal is to obtain cadenced observations of brighter, moderate redshift ($z \simeq$ 2--3) transients while allowing for deep yearly coadds for the highest redshift transients.

\item \textbf{We also recommend re-observation of the COSMOS-Web extragalactic deep field} \citep{Casey:2023}. This equatorial wide-area survey provides an unmatched pre-existing JWST ``template'' observation for transient discoveries and extensive characterization (including redshifts) of potential host galaxies. We recommend two epochs of NIRCam imaging over one year in F115W, F150W, F277W, and F444W, reaching a 5$\sigma$ depth of 28 mag. An alternative would be to observe in F150W, F277W, and F444W only, going deeper in F150W to help rule out lower-redshift transients. The observations would ideally be scheduled early in the twice-yearly observing window (i.e., November and April) to allow for follow-up of transient discoveries, except the low ecliptic latitude of this field leads to a high and detrimental background in November, so observations in December and April are likely most optimal.

We recommend approximately 100 hours for these observations. This allocation will not allow for complete coverage of 0.54 deg$^2$ COSMOS-Web field, so we recommend prioritizing approximately 0.1 deg$^2$ with the best existing coverage (i.e., including MIRI, NIRSpec, etc.).

Because of the limited JWST observability of this field, the key science goal with these observations will be higher-redshift transients. The already existing JWST observations plus those recommended here should provide exciting high-redshift transients for individual follow-up (see below). This field could be re-observed yearly (or every other year) into the future, but that would require an additional allocation of observing time beyond the budget of our current recommendation. This option should be considered.

\item The highest redshift transients discovered by JWST may be found in a lensing cluster field, enhancing JWST’s reach with a cosmic telescope. \textbf{We recommend approximately twice-yearly observations for three or more years of one or more cluster lenses that are the most favorable for revealing very high redshift transients.} NIRCam observations in F150W, F277W, and F444W are recommended, with a total of approximately 40 hours for this part of the program (with the tradeoff between the number of fields and number of epochs to be determined).

\item \textbf{At high priority we recommend approximately 100 hours be reserved for follow-up of exciting high-redshift transients.} These do not necessarily need to be sources discovered from this program; high-redshift transients discovered at other wavelengths (e.g., GRBs), with other facilities, or in other fields, that would benefit from JWST follow-up could also be observed using this pool. Any kind of useful follow-up should be considered, including higher cadence imaging to get light curves, different imaging filters, or spectroscopy. 

Requests to use this follow-up pool of time should be open to all. We recommend a standing committee to help review these requests, with membership including STScI staff as well as representatives from the broader transient community. This would allow observing requests to be reviewed in a more holistic fashion than the current ad hoc reviews for DDT proposals. The committee would be able to more consistently judge the relative importance of different objects and follow-up observations, as well as keep a big-picture view of the range of object types and science. The committee should adopt a transparent review process and seek community engagement with open communication.

This community process and reserved pool should not preclude DDT observation requests made through the normal channel nor should it preclude GO or other types of proposals that include JWST followup of high-redshift transients. \textbf{Indeed, we encourage the design of this program to enable additional GO proposals for related and ancillary science.} We do recommend, however, that GO observations of these fields (in additional filters, at other epochs, etc.) \textit{that leverage existing public data from this program and that would be useful for further transient discovery or characterization} waive any proprietary period.

\item Parallel imaging to the main surveys will be incredibly valuable. \textbf{We recommend, at a minimum, the NIRCam imaging should be supplemented with one of NIRISS imaging, NIRSpec spectroscopy, or potentially MIRI imaging, in parallel.} The mode of parallel observations should be determined from their scientific value as well as observational considerations to maximize the sky area repeated in the prime and parallel fields so they are most useful for detection and monitoring of transients.

For the CVZ imaging, we anticipate NIRISS imaging will be the most valuable addition (increasing sky area for transient discovery), whereas for COSMOS-Web the NIRCam imaging observations should be arranged to allow for parallel NIRSpec spectroscopy of (likely thousands of) high-redshift sources identified in existing data (and prioritized through a community process). The lensing cluster NIRCam imaging could be supplemented by either NIRISS imaging or NIRSpec spectroscopy.

However, \textbf{we also recommend STScI to explore using more than two instruments in parallel for this program, including the possibility of science data from FGS.} The transient yield of this survey scales directly as the sky area observed; the more instruments observing in parallel, the better.

\item \textbf{HST observations can also be valuable to this survey.} We especially recommend HST UV and optical coverage of the CVZ survey fields to enable galaxy science and photometric redshifts that will be needed to vet JWST transient discoveries. HST should also be used in followup of discovered transients (e.g., to help discriminate lower-redshift transients).

\item \textbf{We recommend establishing an implementation team at STScI for the survey.} The observing plan should be broadly announced and the high-redshift transient community should be encouraged to share discoveries, follow-up plans, etc. STScI should also allocate resources (people, computation, etc.) to create infrastructure to interface with the survey, including data searching, storage, and delivery of high-level science products (images, coadds, subtractions, transient alerts, catalogs, etc.) to the community in a structured and easy-access way. Other public HST and JWST data on these fields should be incorporated and presented in a uniform way. This will allow broad access and provide a platform for the community to build upon these data, e.g., by conducting follow-up multiwavelength campaigns with other facilities.

\end{enumerate}

\clearpage

\section*{Acknowledgments}

We are grateful to all of the astronomers who provided input to our working group. We thank Nancy Levenson for the opportunity to contribute to these important topics. We also express our appreciation to Laura Watkins and Neill Reid for all of their assistance, as well as to the HST and JWST User Committees for input and feedback. We thank Jane Rigby for alerting us to the significant seasonal background variation in the COSMOS field.

\bibliography{ltvm}

\clearpage 

\appendix

\section*{Appendices}

\section{Call for Input \label{sec:call_for_input}}

Email sent by STScI on June 15, 2023:

Dear colleagues,

The HST/JWST Long-Term Monitoring Working Group is soliciting community input to develop recommendations to take best advantage of long time baseline observations spanning three decades of past HST observations and looking forward to two decades of JWST data. More information about our working group can be found in our charter.

We encourage input either by completing this short, on-line survey and/or in the form of short contributions submitted to STScI by Friday, September 8, 2023. We request these be in PDF format and ideally limited to 1 page (+ figures/references), but any reasonable length will be accepted. Contributions do not need to be anonymized and multiple co-authors are welcome. Submissions will guide the working group recommendations, but will not be shared publicly. 

PDF contributions should be e-mailed to wg-longterm@stsci.edu by the September 8th deadline.

Submissions should explicitly specify which of these two topics is being addressed:

\begin{itemize}

\item {\bf Long time baseline science opportunities:} We are interested in learning about the key science that would be enabled with long time baseline observations and science themes that should be prioritized in General Observer and Archival proposals. For our purposes, long time baselines refer to those that are not easily accommodated in the standard proposal process. Observations could include, but are not limited to, photometric or spectroscopic variability, and astrometric motions. 
 
\item {\bf JWST DDT for high-redshift transients:} Our working group has also been directed to develop a specific concept for a Director’s Discretionary Time observing program that will use JWST’s imaging and spectroscopic capabilities to probe transient phenomena at high redshift, with a goal of starting implementation of the program in JWST Cycle 2. We are interested in understanding the science cases that should be prioritized for such a program.

\end{itemize}

In all cases, in addition to the science, we are soliciting input about the observational resources that would be required, advice about timing and sky location for the observations, and suggestions for mechanisms to promote this science in the broad portfolio of HST and JWST programs.

The working group is expected to develop preliminary recommendations based on the received input in the (northern) fall of 2023, with final recommendations in early 2024.

Thank you, \\
The HST/JWST Long-Term Monitoring Working Group

\section{Community Survey \label{sec:community_survey}}

The HST/JWST Long-Term Monitoring Working Group is soliciting community input to develop recommendations to take best advantage of long time baseline observations spanning three decades of past HST observations and looking forward to two decades of JWST data. More information about our working group can be found in our charter.

This survey is designed to gather information for the working group recommendations. All questions are optional; feel free to answer as many or as few as you wish. 

Feedback to the working group can also be submitted in the form of short contributions emailed to STScI as described in the recent email announcement.  

\begin{itemize}

\item Science categories

\begin{itemize}

\item Considering the categories below, please choose three that you would prioritize for long-term monitoring science with HST and JWST.
 
Transients \\
Solar system objects \\
Variable stars (Galactic, extragalactic, long and short variability) \\
AGN \\
Nearby galaxies \\
Proper motion studies (nearby galaxies and clusters) \\
Software/Infrastructure \\

\item Please elaborate on your choices above, if you wish.

\end{itemize}

\item Archival data and future observations

\begin{itemize}

\item What existing data in the HST/JWST archives should be supplemented by future HST/JWST observations to enable long-term science?

\item What JWST/HST observations should be obtained soon to allow for future long-term science with JWST?

\item What JWST/HST observations should be obtained soon to allow for future science with upcoming facilities such as the Nancy Grace Roman Space Telescope, large ground-based telescopes, and/or future space telescopes?

\end{itemize}

\item HST+JWST combined projects

\begin{itemize}

\item List ONE specific science topic you would explore using the combined capabilities of HST and JWST. What kind of observing program would be required?

\end{itemize}

\item JWST DDT for high-redshift transients

Our working group has also been directed to develop a specific concept for a Director’s Discretionary Time observing program that will use JWST’s imaging and spectroscopic capabilities to probe transient phenomena at high redshift, with a goal of starting implementation of the program in JWST Cycle 2.

\begin{itemize}

\item What science questions should such a program aim to address?

\item What observations would be needed to address those questions? 

\end{itemize}

\item Archive tools and proposing

\begin{itemize}

\item What archival tools or capabilities do you see missing from the current MAST that would improve long-term monitoring science?

\item What changes to the proposal process are needed to implement long time baseline observations?

\end{itemize}

\item Additional comments

\begin{itemize}

\item Please provide any additional comments you may have.

\end{itemize}

\end{itemize}

\end{document}